\documentclass[prb,aps,showpacs,preprint]{revtex4-1}
\usepackage{graphicx,color,amsmath,amssymb}
\usepackage{bm,hyperref}

\begin{document}

\title{Generalization of Laughlin's Theory for the Fractional Quantum Hall Effect}
\author{Sudhansu S. Mandal}
\affiliation{Department of Physics and Centre for Theoretical Studies, Indian Institute of Technology Kharagpur, West Bengal 721302, India}




\begin{abstract}
Motivated by the structure of the quasiparticle wavefunction in the composite fermion (CF) theory for fractional quantum Hall filling factor $\nu = 1/m$ ($m$ odd), I consider a suitable quasiparticle operator in differential form, as a modified form of Laughlin's quasiparticle operator, that reproduces quasiparticle wave function as predicted in the CF theory, without {\em a priori} assumption of the presence of CFs. I further consider the conjugate of this operator as quasihole operator for obtaining a novel quasihole wave function for $1/m$ state. 
Each of these wave functions is interpreted as expelled electron into a different Hilbert subspace from the original Hilbert space of Laughlin condensate while still maintaining its correlation (although changed) with the electrons in the condensate such that the expelled electron behaves as a CF with respect to the electrons in the condensate.  
With this interpretation, 
I show that the ground state wavefunctions for general states at filling fractions $\nu_{n,m}^{\pm} = n/[n(m\mp 1)\pm 1]$, respectively, can be constructed 
as coherent superposition of $n$ coupled Laughlin condensates and their ``conjugates'', formed at different Hilbert subspaces.
 The corresponding wave functions, specially surprising for $\nu_{n,m}^{-}$ sequence of states, 
 are identical with those proposed in the theory of noninteracting CFs.
 The states which were considered as fractional quantum Hall effect of interacting CFs, can also be treated in the same footing as for the prominent sequences of states describing as   the 
 coupled condensates among which one is a non-Laughlin condensate in a different Hilbert subspace.
 Further, I predict that the half filling of the lowest Landau level is a  quantum critical point for phase transitions between two topologically distinct phases each corresponding to a family of states: One consists of large number of coupled Laughlin condensates of filling factor $1/3$ and the other corresponds to large number of coupled conjugate Laughlin condensates of filling factor $1$, which may be distinguished, respectively, by the absence and presence of upstream edge modes. 
 
\end{abstract}

\maketitle

\section{Introduction}

Correlated quantum liquids formed by two-dimensional electrons exposed to a large magnetic field manifest through the phenomenon of fractional quantum
Hall effect (FQHE) \cite{Tsui82}, characterized by different topological quantum numbers labeled \cite{Willett87} with the corresponding filling fractions. 
The single particle wave function for two-dimensional electrons in the lowest Landau level (LL)
subjected to a magnetic field $B$ perpendicular to the plane of the system is 
$\psi_l(z) =z^l \exp[-\vert z \vert^2
/4]$ with complex coordinate $z= (x-iy)/\ell_B$, angular momentum of the degenerate orbitals $l=0,1,2,\cdots$, and magnetic length $\ell_B = (\hbar c /eB)^{1/2}$.
The manybody wave function describing any correlated liquid, in principle, will be an appropriate linear combination of wave 
functions for occupied orbitals; however this will not be useful for large number of electrons as we do not have any general procedure 
for extracting any meaningful convenient form of the wave function from it.
Nonetheless 
Laughlin \cite{Laughlin83}, based on certain general principles, proposed very accurate ground state wave function for $N$ electrons 
at the filling fraction $\nu =1/m$   
 (hitherto suppressing ubiquitous Gaussian factor 
for each electron) as $\Psi^L_{1/m} = \prod_{k<l}z_{kl}^m $ with 
$z_{kl} = z_k -z_l$, where $z_k$ is the coordinate of $k$-th
electron. The wave functions of a quasiparticle (QP) and a quasihole (QH) excitations of the Laughlin states are $\Psi_{1/m}^{L,\rm{qp}} = \prod_l
\partial_{z_l} \Psi^L_{1/m}$ and $\Psi_{1/m}^{L,\rm{qh}} = \prod_l z_l \Psi^L_{1/m}$ with their respective charges $(-e/m)$ and $e/m$. The Haldane-Halperin (HH)
hierarchy theory \cite{Haldane83,Halperin84} for other fractions in the lowest LL
predicts the production of daughter states by the condensation of these QPs and QHs at certain densities.
However, no simple \cite{Morf86,Morf87,Greiter94} explicit wave functions for these daughter states can be formulated using QP or QH wave functions.


The composite fermion (CF) theory \cite{Jain89,Jain90,Jain07}, on the other hand, begins with the postulate that every electron associated with an even number ($2s$) of 
quantized vortices forms a CF (denoted by $^{2s}$CF) which becomes
the effective weakly interacting QP for the FQHE. The integer quantum Hall effect \cite{Klitzing80} of these QPs 
with filling factor $\nu^* = n$ will
produce incompressible states at filling factors $ n/(2sn\pm 1)$, where 
the sign $+(-)$ refers to the parallel (antiparallel) direction of effective magnetic field for the CFs with
respect to the applied magnetic field. 
The corresponding hierarchy is robust as the states that have been observed belong to these
sequences and their energy gaps decrease with the increase of both $n$ and $s$. 
This theory naturally predicts the explicit ground state wave functions 
which have been shown to be fairly accurate \cite{Jain97} for the Coulomb interaction in finite systems, and
it reproduces \cite{Jain89} Laughlin wave function for $\nu =1/(2s\pm 1)$  when $n=1$. 
Further, the CF theory suggests that the CFs form gapless Fermi sea \cite{Kalmeyer92,HLR93} at $\nu = 1/(2s)$ that has been rigorously tested 
\cite{Willett93,Kang93,Goldman94} along with 
the presence of the structure \cite{Kukushkin99,Kukushkin09,Majumder11,Rhone11} of effective Landau-like levels.

In spite of exemplary success of the CF theory, it is not yet clear how the CFs emerge as relevant quasiparticles for the FQHE.
Further, why the 
general principles that fetch Laughlin theory do not reproduce other states that are described by the composite fermion theory is an outstanding issue for about last three decades.
Ideally, we would like to have a connection between the CF picture and the Laughlin description which is the main focus of this study.

I propose a QP operator, as a modified form of Laughlin quasiparticle operator, and its conjugate as QH operator which operating on the Laughlin wave function describing a Laughlin condensate \cite{Laughlin83,Girvin87} (LC) at $\nu = 1/m$ in
Hilbert-subspace of the lowest LL, generate a QP and a QH, respectively, by expelling electrons from the LC to the projected Hilbert-subspace of 
the second LL and its conjugate LL (conjugation of the basis states) onto the lowest LL.
I argue that the expelled electrons are CFs with respect to the LC as their associated order of correlation holes either reduces or increases by one in 
comparison to the electrons in LC.
The condensate formed by these expelled electrons into the fraction $1/m$, together with its coupling to
the original condensate, respectively constitute new fractions $2/[2(m-1)+1]$ and $2/[2(m+1)-1]$. In general, hierarchically constructed fractions
$\nu_{n,m}^{+}=n/[n(m-1)+1]$ and $\nu_{n,m}^{-}=n/[n(m+1)-1]$, respectively, can be described as $n$ coupled LCs and their conjugates of filling fraction $1/m$ each formed at the 
different Hilbert sub-spaces of analytic functions \cite{Girvin84};
the corresponding wave functions are found to be identical with
those proposed in the noninteracting CF theory. The wave functions for the states \cite{Pan03,Pan15,Csathy15} in the range $1/3 <\nu < 2/5$, such as 4/11 and 5/13,
can be described by the two coupled condensates among which one is a non-Laughlin condensate formed in a different Hilbert subspace.  
The present generalization predicts the filling fractions $1/(2s)$ to be the quantum critical points for phase transitions between two distinct  families of topological phases.

Rest of the paper is organised as follows. In Section-II, I describe a quasiparticle and a novel quasihole operators which acting on Laughlin states produce quasiparticle and quasihole wave functions. Both these wave functions are energetically favourable in comparison to Laughlin's quasiparticle and quasihole wave functions. By exploiting the steps of forming more and more quasiparticles, the ground state wave functions for other conventional non-Laughlin FQHE states have been constructed in Section-III. These are coherently coupled either Laughlin condensates or their conjugates. The wave functions for few other states have also been constructed. Possible topological phase transitions at the even denominator filling factors have been discussed in Section-IV. A comparative study between the present and  other hierarchical constructions in the literature has been discussed in section-V. Section-VI is devoted for conclusion.

\section{Quasiparticles and Quasiholes of Laughlin Condensates}

\begin{figure}[h]
\includegraphics[width=10cm]{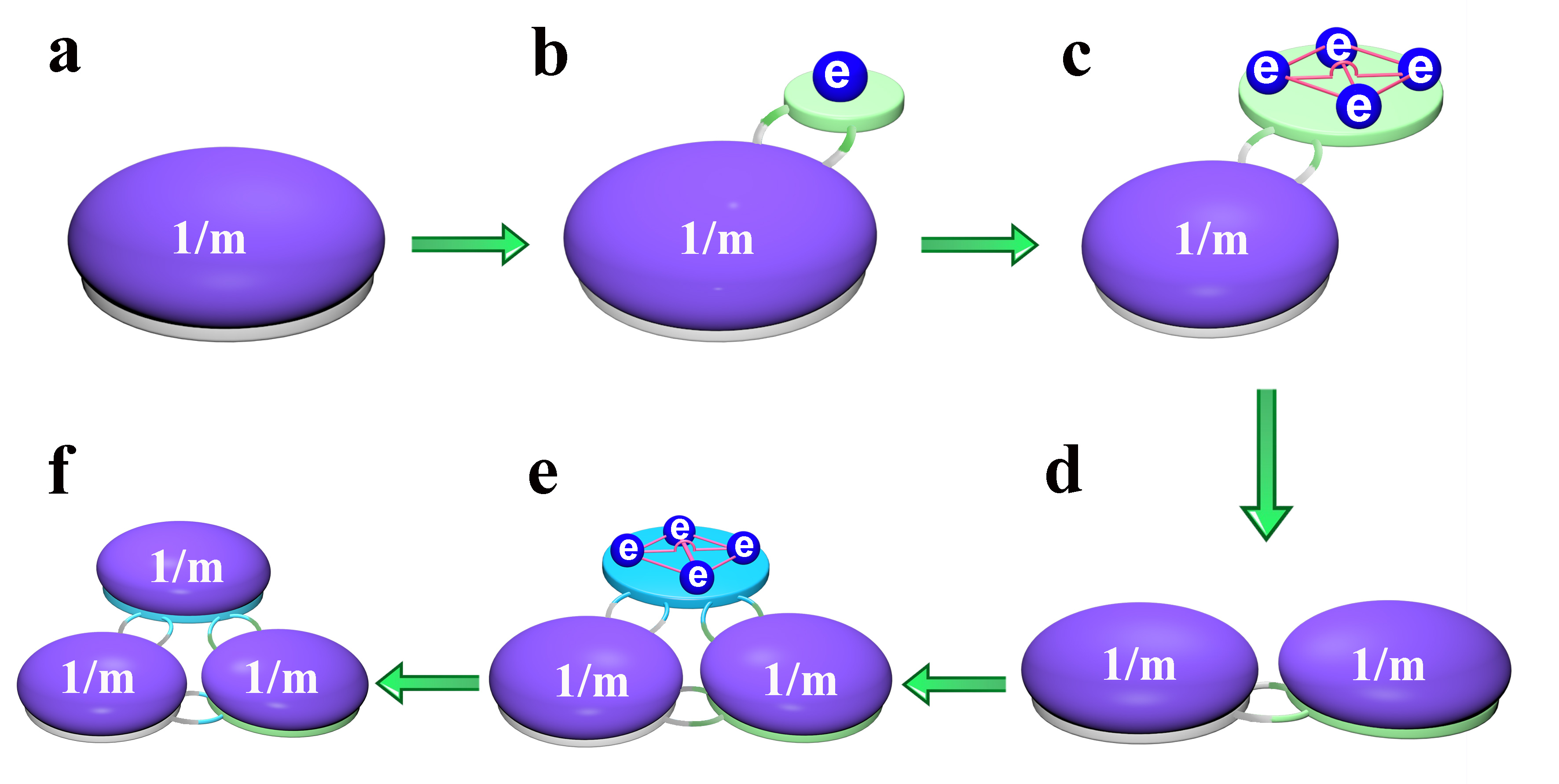}
\caption{(colour online) Schematic diagrams for the steps of transformations on decreasing flux from a Laughlin condensate.
The disk-shaped bases represent different Hilbert subspaces, ellipsoids on these bases represent LCs, lines connected to electrons (spheres) represent strong interaction,
and ring-shaped structures represent coupling between the layers. The arrows represent the evolution of the condensates on decreasing flux.
}
\label{condensates}
\end{figure}

Motivated by the form of the quasiparticle wavefunction \cite{Jain07,Jeon03} within the composite fermion theory  at $\nu =1/m$, I begin with proposing a modified Laughlin-like QP operator in a nonlocal form:
\begin{equation} 
\hat{{\cal O}}_{{\rm qp}} = \sum_{j=1}^N  \left( \frac{\partial}{\partial z_j } \prod_{l\neq j} \frac{\partial}{\partial z_{lj} }\right) 
\label{op_qp}
\end{equation}
This QP operator reduces angular momentum of a selected electron by one
unit and relative angular momentum of other electrons with respect to the selected electron also by one unit. 
While the Laughlin QP operator is the product of the derivatives with respect to all the particle coordinates, this
operator is the sum of the products of derivatives with respect to a selected particle coordinate and relative coordinates of other particles. 
The action of this QP operator on the Laughlin state at $\nu=1/m$ is to produce a state with one reduced flux quantum from the LC (Fig.~\ref{condensates}${\bf a}$). 
A quasiparticle wave function is thus found to be
\begin{eqnarray}
 \Psi^{\rm{qp}}_{1/m} &=& \hat{{\cal O}}_{{\rm qp}} \Psi^L_{1/m}  = \sum_{j=1}^N (-1)^j \frac{\partial}{\partial z_j} T_j^{m-1} \prod^\prime_{i<k} z_{ik}^m  \nonumber \\
 &\equiv& \sum_{j=1}^N (-1)^j P_j T_j^{m-1} \Psi_{1/m}^{(j)} \, ,
 \label{wfn_qp}
\end{eqnarray}
where $\prime$ denotes the exclusion of $j^{{\rm th}}$ electron in the product, and
$\Psi_{1/m}^{(j)} = \prod_{k<l}' z_{kl}^m$ represent LC without $j$-th electron with 
 $P_j = \sum_{k\neq j} z_{jk}^{-1}$, and   $T_j^{m-1} = \prod_{k\neq j} z_{jk}^{m-1}$.
The physical description of this QP in Eq.~\ref{wfn_qp} is as follows:
One of the electrons gets expelled (Fig.~\ref{condensates}${\bf b}$) from the LC where remaining $(N-1)$ electrons rearrange themselves in their Hilbert subspace. 
While each of the electrons in 
the condensate experiences
correlation holes of order $m$ at the position of other electrons, they feel correlation holes of order $(m-1)$ associated with the expelled electron; thus a $^{m-1}$CF 
emerges as QP excitation of the LC and the net increase of charge is $(-e/m)$ due to the reduction of correlation hole by order one.
Although the expelled electron in Eq.~(\ref{wfn_qp}) is shown to be in the zero angular momentum state, it can acquire states with
any angular momentum without costing any extra energy.
Exiting from the LC in the Hilbert subspace, ${\cal H}_1$, characterized by the set of analytic functions
$\{ z_j^\ell\}$, $(\ell = 0,1,2,\cdots$ representing angular momentum of a single-particle state), the electron occupies one of the states 
in the Hilbert subspace, ${\cal H}_2$, characterized by a set of analytic functions \cite{Girvin84} $\{ z_j^{\ell} P_j \}$. 
While the former subspace corresponds to the lowest LL, the latter may be obtained by projecting \cite{Note}
the second LL onto the lowest one for a fixed number of electrons.  
Although these analytic functions may appear to be singular when two electrons approach each other, the QP wave function is not singular due to the coupling of 
expelled electron with the condensate in the form of Jastrow factor $T_j^{m-1}$, and ${\cal H}_2$ exists only when some of the electrons occupy ${\cal H}_1$.
The above functional form of $P_j$ is fixed by the requirement of maximum possible exponent of any coordinate equal to the total number of flux quanta.
The QP wave function in Eq.~(\ref{wfn_qp}) differs from the 
Laughlin QP wave function but is identical with the QP wave function in the CF theory, and thus, as shown in Ref.~\onlinecite{Jeon03},
the energy of $\Psi_{1/m}^{{\rm qp}}$ is lower than that of $\Psi_{1/m}^{L,{\rm{qp}}}$. The form (\ref{wfn_qp}) of quasiparticle wave function can be exploited to construct (see section III-A below) the ground state wave function for FQHE state at the filling factor $\nu_{2,m}^+ = 2/[2(m-1)+1]$.

Once we choose a QP operator, we loose freedom of choosing a QH operator and vice versa. 
The QH operator is then conjugate to the QP operator (\ref{op_qp}),
\begin{equation}
\hat{{\cal O}}_{{\rm qh}} = \sum_{j=1}^N  \left( z_j  \prod_{l\neq j}{z_{lj}} \right)
\label{op_qh}
\end{equation}
which acting on the Laughlin wave function produces a QH wave function
\begin{equation}
\Psi^{\rm{qh}}_{1/m} = \hat{{\cal O}}_{{\rm qh}}\Psi^L_{1/m} \equiv \sum_{j=1}^N   z_j  \prod_{l\neq j}{z_{lj}} \prod_{i<k} z_{ik}^m
\end{equation}
which may be rewritten as 
\begin{equation}
\Psi^{{\rm qh}}_{1/m} = \sum_{j=1}^N z_j \prod_{l\neq j}{z_{lj}}  \prod_{i<k} z_{ik}^{m+1} \left\vert \begin{array}{cccc} 
1 & 1 & \cdots & 1 \\
P_1^{(1)} & P_2^{(1)} & \cdots & P_N^{(1)} \\
P_1^{(2)} & P_2^{(2)} & \cdots & P_N^{(2)} \\
\vdots & \vdots & \ddots & \vdots \\
P_1^{(N-1)} & P_2^{(N-1)} & \cdots & P_N^{(N-1)}
\end{array} \right\vert
\label{qh2}
\end{equation}
as the determinant in Eq.~(\ref{qh2}) is equivalent to (up to a constant factor) the inverse of  the Vandermonde matrix, {\it i.e.}, $\prod_{l<n}z_{ln}^{-1}$ (see supplementary material). Here
\begin{eqnarray}
P^{(n)}_j &=& \sum_{k_1\neq k_2\cdots \neq k_n\neq j} \frac{1}{z_{jk_1}\cdots z_{jk_n}} \\
&=& \sum_{l=1}^n (-1)^{l+1} P_{j,l} P_j^{(n-l)} \frac{(n-1)!}{(n-l)!}
\end{eqnarray}
with $P_{j,l} = \sum_{k\neq j} (\frac{1}{z_{jk}})^l$ and $P_{j,1} = P_j^{(1)}\equiv P_j$.
The quasihole wave function (\ref{qh2}) is then may be re-expressed as
\begin{equation}
\Psi^{{\rm qh}}_{1/m} = \sum_{j=1}^N (-1)^j z_j T_j^{m+1} \tilde{\Psi}_{1/m}^{(j)} \, ,
\label{wfn_qh}
\end{equation}
because $\prod_{l\neq j} z_{lj}^{-1} = P_j^{(N-1)}$ and $T_j^{m+1} = \prod_{l\neq j} z_{lj}^{m+1}$. The wave function of a ``conjugate'' Laughlin condensate \cite{SM} (CLC) without $j^{\rm th}$ electron is given by
\begin{widetext}
	\begin{equation}
	\tilde{\Psi}_{1/m}^{(j)} =  \prod_{i<k}' z_{ik}^{m+1} \left\vert   
	\begin{array}{ccccccc}
	1 & 1 & \cdots & 1 & 1 & \cdots & 1 \\
	P_1^{(1)} & P_2^{(1)} & \cdots & P_{j-1}^{(1)} & P_{j+1}^{(1)} & \cdots & P_N^{(1)}  \\
	P_1^{(2)} & P_2^{(2)} & \cdots & P_{j-1}^{(2)} & P_{j+1}^{(2)} & \cdots & P_N^{(2)}  \\
	\vdots & \vdots & \cdots & \vdots & \vdots & \cdots & \vdots \\
	P_1^{(N-2)} & P_2^{(N-2)} & \cdots & P_{j-1}^{(N-2)} & P_{j+1}^{(N-2)} & \cdots & P_N^{(N-2)}
	\end{array}
	\right\vert 
	\end{equation}
\end{widetext}
where $\prime$ represents the exclusion of $j^{{\rm th}}$ particle in the product. The form (\ref{wfn_qh}) of quasihole wave function is novel and rich.
The wave functions (\ref{wfn_qp} and \ref{wfn_qh}) represent excitations due to decrease and increase of one unit flux quantum from the LC at $\nu = 1/m$, respectively.
The quasihole wave function (\ref{wfn_qh}) can thus also be represented as an expelled electron from the Laughlin condensate, akin to the quasiparticle representation above, and the resultant quasihole wave function becomes suitable starting point for constructing (Section III B) ground state wave function  which will be exactly same as the CF wave function  at $\nu_{2,m}^- = 2/[2(m+1)-1] $.

\begin{figure}[h]
	\includegraphics[width=7cm]{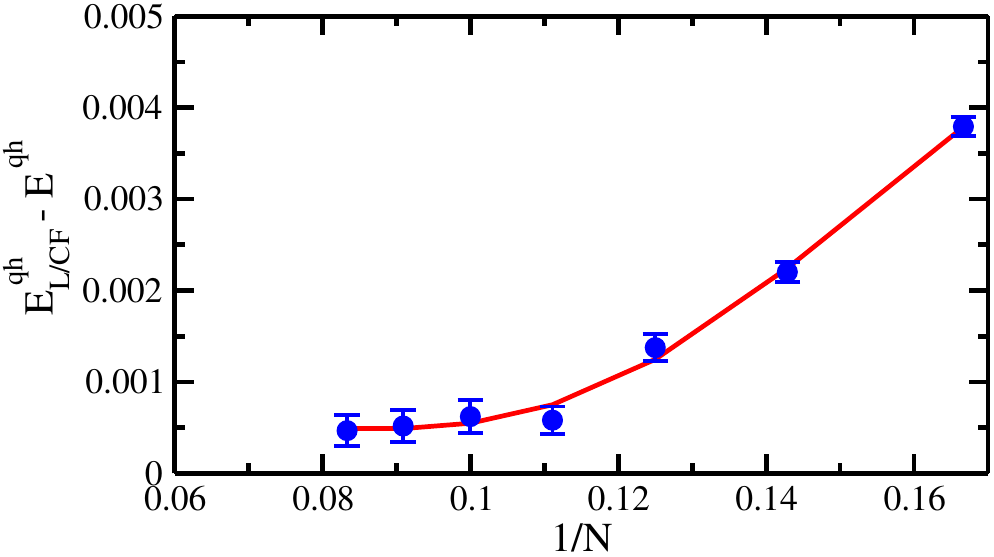}
	\caption{(Colour online) Difference of Coulomb energies (in the unit of $e^2/ \epsilon l_B)$ between Laughlin or CF quasiholes and the proposed quasihole (\ref{wfn_qh}) 
		at $\nu = 1/3$ calculated for different $N$. Here $\epsilon$ is the dielectric constant of the background of the two-dimensional electron gas and $l_B$
		is the magnetic length.}
	\label{energy_quasihole}
\end{figure}

The wave function of a LC at $\nu = 1/m$ is identical (see supplementary material) with the wave function of a CLC at $\nu = 1/m$ when all the electrons are present in the condensate.
We thus can interpret the condensate at $\nu = 1/m$ also as a CLC produced in the Hilbert subspace ${\cal H}_1^\ast$ characterized by a set of additional
analytic functions $\{ P_j^{(\ell)} \}$.
The physical description of a QH is similar to the description of a QP:  
Upon increasing one unit of flux at $\nu = 1/m$, one of the electrons gets expelled from the condensate to the Hilbert subspace ${\cal H}_2^\ast$ characterized by
a set of additional analytic functions $\{ z_jP_j^{(\ell)} \}$. 
Since the expelled electron is associated with correlation holes of order $(m+1)$ with respect to all other electrons in the CLC where every electron feels
correlation holes of order $m$ in the positions of other electrons, the expelled electron can be considered as an emergent $^{m+1}$CF from the condensate and 
the excess of one order in correlation
hole for the expelled electron justifies the QH with charge $e/m$. 
The QH wave function in Eq.~(\ref{wfn_qh}) represents a QH with zero angular momentum, but in principle, QH may have any angular momentum without costing any extra energy.
While the QH wave wave functions for the Laughlin and the CF theories are identical,
the QH wave function here is different and has lower energy (Fig.\ref{energy_quasihole}) than the energy of former two QHs.

\begin{figure}[h]
	\includegraphics[width=6cm]{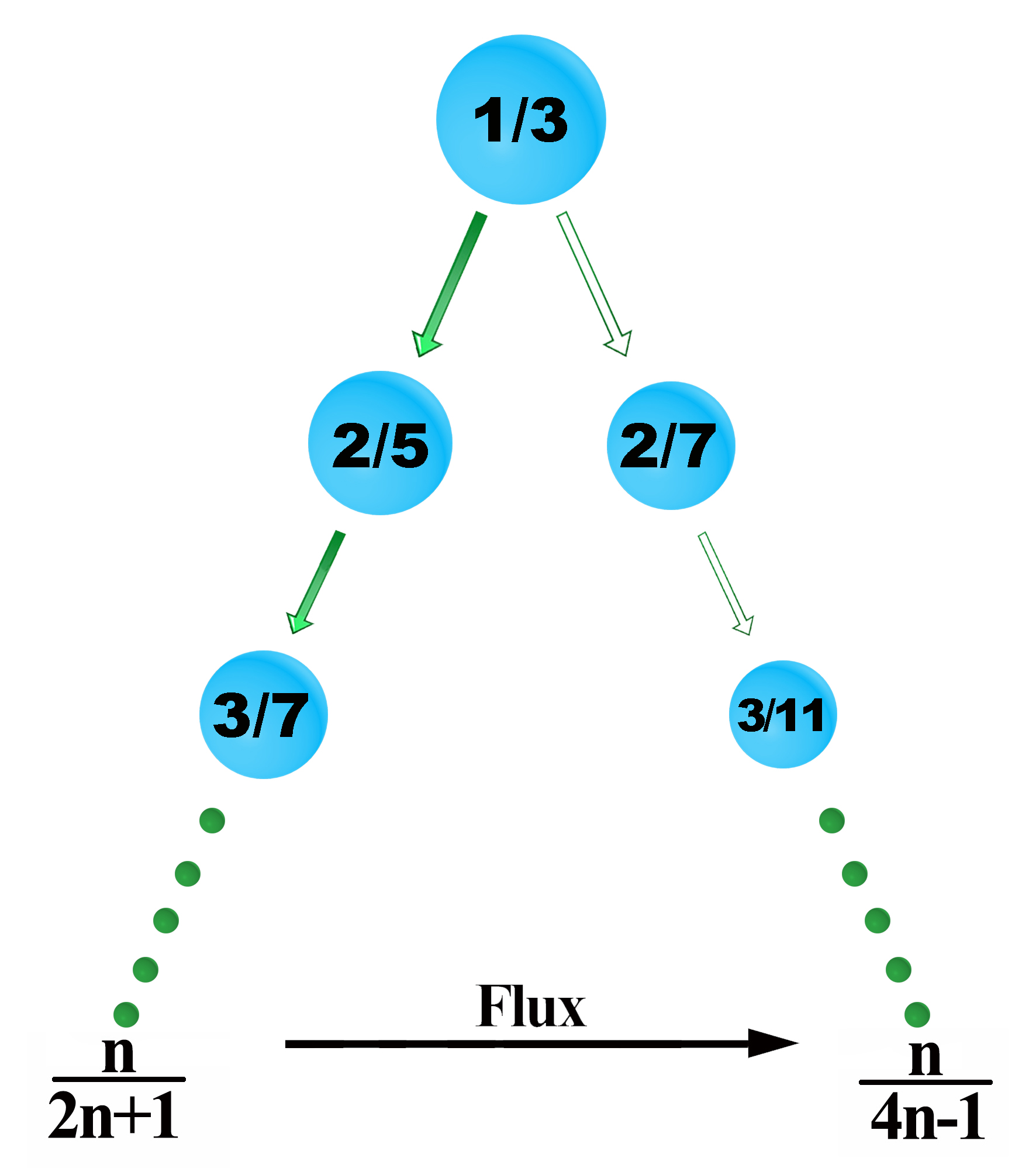}
	\caption{(Colour online) Both the sequences of filling factors $\frac{n}{2n+1}$ and $\frac{n}{4n-1}$ starting at $\nu = 1/3$ form two legs. First few states of
		the sequences are shown and the ellipsis represent the continuation of the sequences. The direction of the increment of flux in the diagram 
		is shown by an arrow. The arrows connecting two consecutive states in the sequences represent whether the states are generated by the 
		QP (filled arrows) or QH (unfilled arrows) excitations of their predecessors.}
	\label{hierarchy}
\end{figure}

Note that a QP (QH) is represented by the generation of a $^{m-1}$CF ($^{m+1}$CF) out of a LC at $\nu = 1/m$. This may apparently look like an inconsistency, but
it is justified with continuity as follows. While the states below $1/m$ are represented by the sequence $\nu_{n,m}^{+}$ for $^{m-1}$CFs,
the states above $1/m$ are represented by the sequence $\nu_{n,m}^{-}$ for $^{m+1}$CFs; the state at $\nu = 1/m$ is a dichotomy as it belongs to both the sequences
that are constituted with different types of CFs. This has been elucidated in Fig.~\ref{hierarchy} for $m=3$. In the next section, we construct ground state wave functions for $\nu_{n,m}^{+}$ states as $n$ coupled LCs of $1/m$ filling formed at different Hilbert subspaces and for $\nu_{n,m}^{-}$ states as $n$ coupled CLCs of $1/m$ filling. The weight factor of the LCs or the CLCs as the case may be will depend on the prefactors of the analytic functions for that particular Hilbert subspace and interconnectivity of the Hilbert spaces will be given by $T_j^{m-1}$ for $\nu_{n,m}^{+}$ states and $T_j^{m+1}$ for $\nu_{n,m}^{-}$ states, as elucidated for ${\cal H}_2$ and ${\cal H}_2^\ast$, respectively, in Eqs. (\ref{wfn_qp}) and (\ref{wfn_qh}).

\section{Construction of the non-Laughlin States from a Laughlin State}

\subsection{States on Reducing Magnetic Flux}

Decrease of
$q$ flux quanta from the Laughlin ground state leads to a LC with $(N-q)$ electrons in ${\cal H}_1$ and the remaining $q$ electrons 
occupying ${\cal H}_2$ strongly interact (Fig.\ref{condensates}{\bf c}).
The electrons in ${\cal H}_2$ are the composite fermions with respect to the electrons in ${\cal H}_1$ and {\em vice versa}, and these two species of electrons are correlated through $(m-1)$ correlation holes.
In each step of the decrement of quantum of flux from its commensurate value for forming a LC for all electrons, the maximum exponent of each electron coordinate in the reduced LC is equal to the number of flux. 
The decrease of one more unit of
flux quantum after $N/2$ electrons get expelled into ${\cal H}_2$, electrons will not be transfered into ${\cal H}_2$ from ${\cal H}_1$. Instead, this decrease of flux helps to form a LC for the electrons in ${\cal H}_2$, as the flux becomes commensurate for the required number of zeros to form the new LC.
The commensurate flux here refers to which the maximum exponent of all the electrons in the polynomial of the corresponding wavefunction, no matter  which Hilbert subspace they are situated in, is equal to the total number of flux.
The coupling of these two LCs through their associated correlation holes form a topologically distinct condensate (Fig.\ref{condensates}{\bf d}) 
at the filling factor $\nu_{2,m}^{+}$.
Based on this description which follows from the quasiparticle wave function (\ref{wfn_qp}), the explicit form of the constructed ground state wave function for $\nu_{2,m}^+$ state thus can be written as
\begin{eqnarray}
 \Psi_{\nu_{2,m}^+} &=& \sum_{j_1<\cdots <j_{N/2}} \left[ \prod_{j\in \{j_i\}} (-1)^j \prod_{l\notin \{j_i\}}z_{jl}^{m-1} \right] \nonumber \\ 
 &\times& \Psi_{1/m}^{[2]}(\{z_j\}, j\in \{j_i\})
 \,     \Psi_{1/m}^{[1]}(\{z_l\}, l\notin \{j_i\})
 \label{Wave_25}
\end{eqnarray}
which is precisely \cite{SM} the Jain wave function \cite{Jain89,Jain07,Jain97} when $2s = m-1$. The condensates of ${\cal H}_1$ and ${\cal H}_2$ are 
related via $\Psi_{1/m}^{[2]}(\{z_j\}) =
\left( \prod_{j\in \{ j_i\}} P_j \right) \Psi_{1/m}^{[1]} (\{z_j\})$ as ${\cal H}_1$ and ${\cal H}_2$, respectively, are spanned by the functions $\{z_j^\ell\}$ and $\{ P_jz_j^\ell \}$.
The wave function in Eq.~(\ref{Wave_25}) is analogous to the Halperin wave function \cite{Halperin83} $(m,m,m-1)$ for a bilayer system;
the layers here correspond to different Hilbert subspaces.


A further decrease of one unit of flux quantum right at the filling factors $\nu_{2,m}^+$ will
 force to exit one electron each from the two LCs in ${\cal H}_1$ and
${\cal H}_2$ to ${\cal H}_3$, characterized by a set analytic functions $\{z_j^\ell P_j^{(2)} \}$ 
where $P_j^{(l)} = \sum_{k_1,\cdots,k_l\neq j} \left( z_{k_1j}^{-1}\cdots z_{k_lj}^{-1} \right)$, 
but remains coupled with both the condensates through the correlation holes. 
Decrease of one unit of flux will thus create excitation of charge $(-\nu_{2,m}^+ e)$. 
If we decrease $q$ flux quanta from the flux which creates the condensate at $\nu_{2,m}^+$, 
the LCs of ${\cal H}_1$ and ${\cal H}_2$ will contain $(N/2 -q)$ electrons each and $2q$ electrons in ${\cal H}_3$ will strongly interact (Fig.\ref{condensates}{\bf e}). 
Decrease of $(N/6+1)$ quanta of 
flux will create three LCs of $\nu = 1/m$ in ${\cal H}_1$, ${\cal H}_2$ and ${\cal H}_3$ with $N/3$ number of electrons each and their mutual coupling will manifest 
(Fig.\ref{condensates}{\bf f}) a new condensate \cite{SM} at the filling 
factor $\nu_{3,m}^+$. The wave function for this condensate is then can be constructed as
\begin{widetext}
	\begin{eqnarray}
	\Psi_{\nu_{3,m}^+} &=& \sum_{k_1<\cdots <k_{N/3}} \sum_{j_1<\cdots <j_{N/3}}^{\notin \{k_i\}} (-1)^{\sum_j n_j}
	\left[ \prod_{k \in \{k_i\}} (-1)^k \prod_{j \in \{j_i\}} (-1)^j \prod_{l \notin
		\{k_i\}, \{j_i\}} 
	z_{jk}^{m-1} z_{jl}^{m-1} z_{kl}^{m-1} \right] \nonumber \\
	& & \times \Psi^{[3]}_{\frac{1}{m}}(\{z_k\}, k\in \{ k_i \}) \,
	\Psi^{[2]}_{\frac{1}{m}}(\{z_j\}, j\in \{ j_i \})
	\,  \Psi^{[1]}_{\frac{1}{m}} (\{ z_l \}, l \notin \{\{ j_i \},\{k_i\}\})
	\label{wfn_nu3+}
	\end{eqnarray}
\end{widetext}
having angular momentum $M_3^+ = (N/2)[N/\nu_{3,m}^+ - (m+2)]$, where the LC in the Hilbert subspace ${\cal H}_n$ relates with the LC in ${\cal H}_1$  as
$\Psi^{[n]}_{1/m}(\{ z_j\}) = (\prod_j P_{j\in \{ j_i \}}^{(n-1)})\Psi^{[1]}(\{z_j \})$ and $n_j$ is the number of elements in the set $\{k_i\}$ greater than $j$ in the 
set $\{j_i\}$.

To be general, any state in the sequence $\nu_{n,m}^+ = n/[n(m-1)+1]$  will consist of mutually coupled $n$ Laughlin $1/m$ condensates formed in ${\cal H}_1,
\cdots$, and ${\cal H}_n$ with $N/n$ electrons in each. 
Decrease of one unit of flux quantum will expel $n$ electrons (one from each LC) into ${\cal H}_{n+1}$ creating QPs with total
charge $(-\nu_{n,m}^+ e)$. A decrease of $N/[n(n+1)]+1$ units of flux quanta 
will create coupled $1/m$ LCs formed in ${\cal H}_1,\cdots $, and ${\cal H}_{n+1}$  and the system moves to a new topologically distinct state at the filling factor $
\nu_{n+1,m}^+=(n+1)/[(n+1)(m-1)+1]$ .  
The ground state wave functions $\Psi_{\nu_{n,m}^+}$ at the filling factors $\nu_{n,m}^+$ are all identical with those proposed in the noninteracting CF theory.
The corresponding sequence for $m=3$ given by $\nu = \frac{n}{2n+1}$ (see Fig.~\ref{hierarchy}) is the same sequence predicted by the theory of no-interacting $^2$CFs. All the states in this sequence are constructed using coupled 1/3 LCs formed in different Hilbert subspaces.

\subsubsection{States Between Two Consecutive $\nu_{n,m}^+$ States}

The present description not only precisely reproduces the wave functions for the states described by noninteracting CFs, it can easily construct the 
wave functions for the states that can otherwise be understood through the model of CFs when their interactions are taken into account.
The basic principle for forming any incompressible state in the range $\nu_{n,m}^+< \nu < \nu_{n+1,m}^+$ 
is the condensation of interacting electrons in ${\cal H}_{n+1}$ into an incompressible liquid and the coupled state of this condensate with $n$ LCs 
formed in ${\cal H}_1,\cdots,{\cal H}_n$ become an eigenstate of total angular momentum.
As for example, when $N/4+2$ flux quanta are reduced from a LC at 1/3 filling, the 
coupling of LC at 1/3 filling with $(3N/4-2)$ electrons in ${\cal H}_1$ 
and an unconventional condensate \cite{Arek04}, $\bar{\Psi}_{1/5}^{[2]}$, characterized by the zero-energy
ground state of Haldane pseudopotential \cite{Haldane83}, $V_3$, with relative angular momentum $3$, at
1/5 filling with $(N/4+2)$ electrons in ${\cal H}_2$ form a condensate at $\nu =4/11$. The corresponding wave function can thus be expressed as 
\begin{eqnarray}
 \Psi_{4/11} &=& \sum_{j_1<\cdots <j_{N/4+2}} \left[ \prod_{j\in \{j_i\}} (-1)^j \prod_{l\notin \{j_i\}}z_{jl}^{2} \right] \nonumber \\ 
 &\times& \bar{\Psi}_{1/5}^{[2]}(\{z_j\}, j\in \{j_i\})
 \,     \Psi_{1/3}^{[1]}(\{z_l\}, l\notin \{j_i\})
 \label{wave_411}
\end{eqnarray}
The angular momentum of this state $M = (N/2)[(11/4)N-5]$ is consistent with the flux-particle relation for incompressible state \cite{Mukherjee14} in spherical geometry.
This wave function is, however, different from the trial wave function \cite{Mukherjee14} of interacting CF theory because the state at 1/5 filling, unlike Laughlin states, cannot be obtained by simple multiplication of the Jastrow factor $\prod_{i<j} z_{ij}^2$ to 
the state at 1/3 filling of $^2$CFs for the pseudopotential $V_3$. Further, energy of the wavefunction (\ref{wave_411}) will be less in comparison to the wavefunction proposed in interacting CF theory \cite{Mukherjee14},  as 1/5 state is the zero-energy 
ground state \cite{Arek04} of $V_3$.
Similarly, the wavefunctions for the incompressible states \cite{Mukherjee14,Mukherjee12} at $\nu=5/13$ and $3/8$ can be constructed \cite{SM} 
by the coupling of a LC at 1/3
filling formed at ${\cal H}_1$ and another condensate formed at ${\cal H}_2$ with filling fractions 2/7 and 1/4, respectively as follows.
 
A decrease of $(2N-2)/5$ units of flux quanta from $\nu = 1/3$ state will create an unconventional condensate of $(2N-2)/5$ electrons in ${\cal H}_2$ at the filling factor
$2/7$. The ground state wave function at $\nu = 5/13$:
	\begin{eqnarray}
	\Psi_{5/13} &=& \sum_{j_1<\cdots<j_{(2N-2)/5}} \left[ \prod_{j\in \{ j_i\}} (-1)^j \prod_{l \notin \{ j_i \}} z_{jl}^{2} \right] \nonumber \\
	&\times& \bar{\Psi}^{[2]}_{\frac{2}{7}}
	(\{z_j\}, j\in \{ j_i \})
	\, \Psi^{[1]}_{\frac{1}{3}} (\{ z_l \}, l \notin \{ j_i \})
	\label{wave_513}                                           
	\end{eqnarray}
can then be described as coherently coupled LC at $\nu = 1/3$ and this unconventional condensate (characterized by $V_3$) at $\nu = 2/7$. The angular momentum of this state 
$M_{5/13} = (N/2)[\frac{13}{5}(N-1)]$ is also consistent with the predicted flux-particle relationship\cite{Mukherjee14}.

A fractional quantum Hall state at $\nu = 3/8$ can then also be described as coherently coupled LC of $(2N/3)$ electrons at $\nu = 1/3$ in ${\cal H}_1$ and a condensate at $\nu = 1/4$ with Anti-Pfaffian pair correlation \cite{Moore91,LHR07} for $N/3$ electrons in ${\cal H}_2$. Therefore the corresponding ground state
wave function
	\begin{eqnarray}
	\Psi_{3/8} &=& \sum_{j_1<\cdots<j_{N/3}} \left[ \prod_{j\in \{ j_i\}} (-1)^j \prod_{l \notin \{ j_i \}} z_{jl}^{2}\right]   \nonumber \\ &\times & \bar{\Psi}^{[2]}_{\frac{1}{4}}
	(\{z_j\}, j\in \{ j_i \})
	\,\Psi^{[1]}_{\frac{1}{3}} (\{ z_l \}, l \notin \{ j_i \})
	\label{wave_38}                                            
	\end{eqnarray}
	has angular momentum $M_{3/8} = (N/2)[8N/3 -3]$ obeying consistent flux-particle relationship\cite{Mukherjee12}.

Although the ground state wave functions (Eqs.\ref{wave_411}--\ref{wave_38}) which are different from the previously proposed \cite{Mukherjee12,Mukherjee14} ones, 
for $\nu = 4/11$, $5/13$, and $3/8$ are proposed here, their validity for the Coulomb interaction has not been checked. The validity of these wave functions will be published elsewhere with the implementation of
the present theory in a spherical geometry and by determining unconventional wave functions at $1/5$, $2/7$, and $1/4$ filling factors, {\it i.e.},  $\bar{\Psi}_{\frac{1}{5}},\, \bar{\Psi}_{\frac{2}{7}}$, and $ \bar{\Psi}_{\frac{1}{4}}.$\\

\subsection{States on Increasing Magnetic Flux}

By the increase of $q$ flux into the Laughlin state at $\nu = 1/m$, the CLC will have $q$ fewer electrons and the $q$ expelled electrons situating in ${\cal H}_2^\ast$
interact among themselves. When $q$ become $N/2$, the CLC will have $N/2$ electrons and the remaining electrons will strongly interact
in ${\cal H}_2^\ast$. These electrons will be benefited to form a CLC, 
because of the commensuration of flux required for a CLC, by the increment of one more unit of flux and the coherent superposition between
these two coupled CLCs will form a condensate at the filling factors $\nu_{2,m}^- = 2/[2(m+1)-1]$.
The explicit wave functions for these states can be constructed, following Eq.(\ref{wfn_qh}), as
\begin{eqnarray}
 \Psi_{\nu_{2,m}^-} &=& \sum_{j_1<\cdots <j_{N/2}} \left[ \prod_{j\in \{j_i\}} (-1)^j \prod_{l\notin \{j_i\}}z_{jl}^{m+1} \right] \nonumber \\ 
 &\times& \tilde{\Psi}_{1/m}^{[2]}(\{z_j\}, j\in \{j_i\})
 \,     \tilde{\Psi}_{1/m}^{[1]}(\{z_l\}, l\notin \{j_i\}) \, .
 \label{Wave_23}
\end{eqnarray}
These are identical \cite{SM} with the Jain wave functions \cite{Jain07,Jain97}. Here the CLCs in ${\cal H}_1^\ast$ and ${\cal H}_2^\ast$ are related 
via $\tilde{\Psi}_{1/m}^{[2]}(\{z_j\})
= \left( \prod_{j\in \{ j_i \}}z_j \right) \tilde{\Psi}_{1/m}^{[1]}(\{z_j\})$.
Similarly, an increase of $(N/6+1)$ flux quanta from the ground state at $\nu_{2,m}^-$ will create three coupled CLCs of $\nu = 1/m$ in the Hilbert subspace
${\cal H}_1^\ast$, ${\cal H}_2^\ast$, and ${\cal H}_3^\ast$ with $N/3$ 
electrons in each and form a new condensate at $\nu_{3,m}^- = \frac{3}{3(m+1)-1}$. The corresponding ground state wave function that can be constructed with  angular momentum $M_3^- 
= (N/2)[N/\nu_{3,m}^- - (m-2)]$ is given by
\begin{widetext}
	\begin{eqnarray}
	\Psi_{\nu_{3,m}^-} &=& \sum_{k_1<\cdots <k_{N/3}} \sum_{j_1<\cdots <j_{N/3}}^{\notin \{k_i\}} (-1)^{\sum_j n_j} \left[ \prod_{k \in \{k_i\}} (-1)^k \prod_{j \in \{j_i\}} (-1)^j \prod_{l \notin
		\{k_i\}, \{j_i\}} 
	z_{jk}^{m+1} z_{jl}^{m+1} z_{kl}^{m+1} \right] \nonumber \\
	& & \times \tilde{\Psi}^{[3]}_{\frac{1}{m}}(\{z_k\}, k\in \{ k_i \})  \,
	\tilde{\Psi}^{[2]}_{\frac{1}{m}}(\{z_j\}, j\in \{ j_i \})
	\,  \tilde{\Psi}^{[1]}_{\frac{1}{m}} (\{ z_l \}, l \notin \{ \{ j_i \},\{k_i\} \})
	\label{wfn_nu3-}
	\end{eqnarray}
\end{widetext}
where the CLC in ${\cal H}_n^\ast$ is related with the CLC in ${\cal H}_1^\ast$  by the relation
$\tilde{\Psi}^{[n]}_{1/m}(\{ z_j\}) = (\prod_j z_{j\in \{ j_i \}}^{n-1})\tilde{\Psi}^{[1]}(\{z_j \})$.
The wave function $\Psi_{\nu_3^-}$ is identical  
with the corresponding CF wave functions \cite{Jain89,Jain97,Jain07}.
Similarly, all the ground state wave functions  $\Psi_{\nu_n^-}$
can be constructed, and all of those will precisely be the CF wave functions.

In general, increase of flux by $N/[n(n+1)]+1$  from the ground state at 
filling factor $\nu_{n,m}^- = n/[n(m+1)-1]$  consisting of
coherent superposition of coupled $n$ CLCs formed in ${\cal H}_1^\ast ,\cdots$, and ${\cal H }_n^\ast$ 
will create a condensate of $(n+1)$ coupled CLCs at filling factor $\nu_{n+1,m}^- = (n+1)/[(n+1)(m+1)-1]$.
All the associated ground state wave functions are identical \cite{SM} with those proposed in noninteracting CF theory.
The sequence $\nu = \frac{n}{4n-1}$ shown in Fig.\ref{hierarchy} arises due to the increase of flux in comparison to the same for Laughlin 1/3 state. Each of the state in this sequence consists of 1/3 filled CLCs which are coupled.

\section{Topological Phase Transitions at Even-Denominetor Fillinigs}

The sequences of filling factors $\nu_{n,m}^+$ which are $n$ coupled LCs of individual filling factors $1/m$, and $\nu_{n,m-2}^{-}$ which are $n$ coupled CLCs of individual filling factors $1/(m-2)$
converge at the filling factor $\nu = 1/(2s)$ when $m= 2s+1$ and $n \to \infty$ . 
Every value of $n$ on either side of $\nu = 1/(2s)$ is characterized by a Chern number. As the filling factors for large $n$ in either side of $\nu = 1/(2s)$ is densely populated, it appears that for a finite range of filling factors on either side of $\nu = 1/(2s)$ realisation of distinct topological phases with definite Chern number should not be possible. However, the present analysis also suggests that there should be another class of topological number to distinguish the entire family of $n/(2sn+1)$ states from  the entire family of $n/(2sn-1)$ states. These two distinct topological phases are possibly characterized by the direction of the edge modes: while all the edge modes for $n/(2sn+1)$ are downstream, upstream modes \cite{Wen1,Wen2} exist for  $n/(2sn-1)$ states.
In addition to $n$-component Chern-Simons gauge theory for distinguishing two consecutive topological phases  \cite{Wen1,Wen2,SM} in any of the sequences, {\it viz.},$  \frac{n}{n(m-1)\pm1}$, there must be another gauge symmetry which distinguishes topological phase of entire family  in the sequence of $\nu_{n,m}^{+}$  from the phase of other family of states in the sequence of $\nu_{n,m-2}^{-}$.

States with $ \nu \lesssim 1/(2s) $ will be topologically distinct from a state with 
$\nu \gtrsim 1/(2s)$ because the former corresponds to the coherent superposition of 
large number of coupled LCs, formed at different ${\cal H}_n$, of $1/(2s+1)$ filling factors and the latter is formed
when the coherent superposition of large number of CLCs, formed at different ${\cal H}_n^*$, of filling factor $1/(2s-1)$.
The bulk energy gap closes at the filling factors $1/(2s)$ which separates above topologically distinct states and thus $\nu_c = 1/(2s)$
serve as the quantum critical points for  topological phase transitions when magnetic flux is tuned around these.
The closing of the bulk gap at $\nu_c$ is also corroborated with the formation of the Fermi sea \cite{Kalmeyer92,HLR93} of $^{2s}$CFs. While the Fermi surface for $\nu < \nu_c$ 
corresponds to the  $^{2s}$CF particles, the $^{2s}$CF holes create the Fermi surface for $\nu > \nu_c$ because attachment of $2s$ vortices 
for these states over-screens the magnetic flux and thus their Hilbert subspace becomes conjugate to the $^{2s}$CF particles. These CF holes of second kind, 
however, should be
differentiated with the literature of the Fermi surface made with the CF holes \cite{Fisher15}
of first kind which are composite fermionized holes (absence of electrons).
The possible connection of this particle-hole symmetry around $\nu = 1/(2s)$ with the proposed Dirac semi-metal \cite{Son15,Viswanath16} of the CFs near $\nu = 1/(2s)$ cannot be ruled out.
While the surface acoustic wave techniques \cite{Halperin97} can probe the length scale, the experiments on photoluminescence spectroscopy
\cite{Goldberg91,Yusa01,Byszewski06,Nomura14} developed for FQHE systems should probe the energy scale for determining the universality class
of the predicted quantum critical point.


\section{Discussion}
 Immediately after proposal of the CF theory, Read \cite{Read90} argued that the HH hierarchy
and the CF states for same $\nu$ are two different descriptions of the same universality class as they both describe same QP charge and braiding
statistics \cite{Arovas84}. On the contrary, Jain \cite{Jain07,Jain14} pointed out that the CF theory is fundamentally different from the hierarchy theory and they cannot be adiabatically
connected as the flux attachment for CFs is nonperturbative. The calculations \cite{Hansson07,Suorsa11} based on the conformal field theory 
show that the CF wave function can be obtained as the 
correlators of certain operators. This construction,
however, is not microscopic and either presumes the existence of the CF theory since the different sets of operators have been chosen representing the 
different effective Landau levels of the CFs or chooses a specific set of parameters out of the zoo of parameter space.
Bonderson \cite{Bonderson12} 
has argued that the CF description of the FQHE is
hierarchical in nature and that the wave functions of a CF state can be constructed using QP wave function of the hierarchically higher CF state, generated 
by the CF theory. This study also explicitly assumes the existence of the CF theory. 
In this paper, I have proposed QP and QH operators which
generate CFs, upon operating on the Laughlin states, without assuming the presence of any CF in the Laughlin states. 
The wave functions here are hierarchically constructed but they are fundamentally different from the HH hierarchy theory: 
(i) While the hierarchically lower states are formed by the condensation of QPs or QHs of its immediate predecessor in HH hierarchy theory, 
all the states here are formed due to the condensation of electrons into LCs or CLCs
in different Hilbert subspaces;
(ii) in contrary to the HH hierarchy theory, all the states in the sequence $n/(4n-1)$ are created by the QH excitations of their immediate predecessors;
(iii) while the state 5/13 may be obtained only by the condensation of QHs created from the parent 2/5 state in HH hierarchy, this state here is 
constructed by the condensation of expelled electrons from 1/3 state into the filling fraction 2/7 in ${\cal H}_2$.



\section{Conclusion}
I have shown that the CFs emerge as quasiparticle and quasihole excitations of LC; the wave functions 
for the filling factors $\nu = n/(2sn\pm 1)$ in the noninteracting CF theory are identical with
the proposed coherent superposition of $n$ coupled Laughlin (conjugate-Laughlin) condensates of filling factors $1/(2s \pm 1)$ 
formed at different Hilbert-subspaces, where an electron
in any of the condensates is felt by all the electrons in other condensates as CFs.
Therefore this work provides an unification of Laughlin and the composite fermion theories,
and I believe that it would put an end to all the differences of opinions regarding hierarchy pictures.
This generalized scheme predicts the ground state wave functions for all (though shown only for 4/11, 5/13, and 3/8)
the fractional quantum Hall states which are in between any two prominent states described above, without considering any higher order effect such as  
FQHE of interacting CFs.
This theory further opens the possibility of future studies exploring 
the universality class of the quantum criticality and possible topological symmetry distinguishing two phases about the critical points at
the filling factors $\nu_c = 1/(2s)$ in the lowest Landau level.



\section*{Acknowledgements}
I thank S. Mukherjee for useful discussions and K. Manna for his help in drawing cartoon diagrams.




\begin{thebibliography}{99}
\bibitem{Note} The single particle eigen states in the second and higher Landau levels have one of the terms as $z_j^\ell (z_j^*)^\lambda $  (ignoring Gaussian factors), where $\lambda = 1, 2,\cdots$ representing higher Landau levels. When we project second and higher Landau levels onto the lowest one we transform $(z_j^*)^\lambda z_j^l \to \partial^\lambda_{z_j} z_j^\ell $ into an operator acting on the wave function for the filled lowest Landau level. This provides the projected second Landau level Hilbert subspace characterized by the additional functions $\{ z_j^\ell P_j^{(\lambda)}\}$ apart from the functions available for projected lower Landau levels, where $P_j^{(1)}=P_j$. See Ref.~\onlinecite{Jain07} for further details of the procedure of projecting effective higher order Landau levels of composite fermions onto the lowest Landau level.
\end{thebibliography}
\end{document}